# SLALOM: a Language for SLA Specification and Monitoring

Anacleto Correia[2,1], Fernando Brito e Abreu[3,1], Vasco Amaral[1]

[1] CITI/FCT/UNL, 2829-516 Caparica, Portugal
[2] IPS/EST, 2910-761 Setúbal, Portugal
[3] DCTI, ISCTE-IUL, 1649-026 Lisboa, Portugal
accorreia@fct.unl.pt, fba@iscte.pt, vasco.amaral@di.fct.unl.pt

**Abstract.** IT services provisioning is usually underpinned by service level agreements (SLAs), aimed at guaranteeing services quality. However, there is a gap between the customer perspective (business oriented) and that of the service provider (implementation oriented) that becomes more evident while defining and monitoring SLAs. This paper proposes a domain specific language (SLA Language for specificatiOn and Monitoring - SLALOM) to bridge the previous gap. The first step in SLALOM creation was factoring out common concepts, by composing the BPMN metamodel with that of the SLA life cycle, as described in ITIL. The derived metamodel expresses the SLALOM abstract syntax model. The second step was to write concrete syntaxes targeting different aims, such as SLA representation in process models. An example of SLALOM's concrete syntax model instantiation for an IT service supported by self-service financial terminals is presented.

## 1. Introduction

Most organizations rely on Information Technology (IT) services to support their business services. IT services are built upon the technical infrastructure (servers and network devices) as well as on systems and application software. Examples of IT services are a corporate email service, an order entry service or those provided to clients of financial institutions by ATMs.

Likewise business services, IT services are nowadays mostly driven by a customer-focused approach [1]. IT service providers usually offer standard service levels or, alternatively, negotiate particular terms by settling a service level agreement (SLA). The process that addresses SLAs definition and monitoring is called Service Level Management (SLM) [1] and is usually part of a broader framework for IT Service Management (ITSM), such as ITIL v3 [2].

In the absence of an SLM process, IT management would be performed by trial and error, leading to over (or under) capacity and inadequate performance and end-users requirements and expectations would be based in desires rather than feasibility or affordability. SLM implementation has several benefits such as the mutual agreement on which are the relevant IT service quality attributes (e.g. availability, performance, and security), the definition of expected service levels (i.e. thresholds) for those quality attributes) and the clarification of the consequences (e.g. penalties for the provider) if service levels are not met. Service level specification and monitoring requires that customer and provider agree (and express unambiguously in the SLA) on the set of metrics for the IT service quality attributes and how will the data required to compute them be collected.

SLA definition and monitoring are open issues [3] in the ITSM domain, mainly due to the following reasons: (1) SLAs are informally specified [4], (2) SLAs specifications are not grounded on process models and (3) SLAs monitoring uses implementation level metrics (e.g. packet collisions, dropped packets, or page faults) instead of using metrics from an higher abstraction level (e.g. service availability, end-to-end response time, or service affordability).

A concept is said to be at a higher abstraction level when its definition and usage is made independently of implementation constraints or specific technological platform. Abstraction should allow us to convey information for different actors (e.g. end-user, service level manager or system administrator) using an adequate representation (e.g. graphical or textual models, with more or less detail).

Working at higher abstraction levels (e.g. models and metamodels) and mapping the resulting levels, was the aim of OMG's initiative named Model Driven Architecture (MDA) [5]. MDA is a framework for software development, where models are pivotal artifacts used for conveying and documenting requirements and design decisions as well as the basis for performing transformations (e.g. to generate executable artifacts). MDA defines four layers, from M0 (instance layer) to M3 (meta-metamodel layer). The metamodel layer (M2) includes the language constructs (aka grammar) used to describe models (M1 layer).

Our objective in this paper is to introduce the abstract and concrete syntax models (both expressed as metamodels, as proposed in [6]) of SLALOM, a domain-specific language (DSL) that is expected to facilitate SLA specification and monitoring. Since, assessing components execution in isolation does not enable the measurement of service quality at a business-level perspective, we intend to use this DSL to support the composition of measurements from a variety of data sources in order to present and justify observed compound measurements (e.g. the contribution of indicators such as dropped packets, page faults, or queries response time, to explain the metric, end-to-end response time of an IT service).

SLALOM's abstract syntax model will restrict the number of valid models to the ones that conform to it, while SLALOM's concrete syntax model will restrict the number of valid models to the ones that have a valid concrete representation [6]. Thus, if process models were chosen to express SLAs contracts, the representation of IT services, should conform to the metamodel that defines the concrete syntax of process models.

In section 2 we will present the DSL that was the source of metamodel composition, as well as the final result: the metamodel that is the abstract syntax model of the SLALOM language. In section 3 the concrete syntax models derived for different purposes are presented: SLAs monitoring and compliance checking SLAs depicted in process models, and validation & verification of properties of SLAs process models. In section 4 we present an illustrative example with concrete syntax metamodels instantiation. In section 5 we overview previous related works regarding SLAs, and finally in section 6 we discuss future work regarding the integration of concrete syntax methods in the process for SLAs specification and monitoring.

## 2. Abstract syntax model

The abstract syntax of a language takes a central position in a language specification since it is the pivot between various concrete syntaxes of the same language, as well as between the syntactical structure of a model and its semantics [6]. The first step in creating the SLALOM language was to identify the concepts to express in its abstract syntax model. For each concept, the semantics was clarified and the relationships with other concepts were elicited.

We have adopted the UML class diagram notation [7] enriched with OCL constraints [8] for abstract syntax model definition. OCL was also used to specify the static semantics of the DSL, that is, the set of rules that specify whether domain models are well formed. Later, this will enable to check concrete syntax models well-formedness against the abstract syntax model.

When we have different DSLs that capture and model a shared set of concepts, those constructs can be joined, by a metamodel composition technique [9], to stitch the two languages together into a unified whole. As such, the new language benefits from previously documented domain knowledge since it reuses, at least partly, the concepts expressed in existing metamodels. We followed this approach to derive the SLALOM abstract syntax model, by composing a metamodel of the SLA life-cycle in the context of SLM process with a BPMN metamodel, as described henceforth.

### 2.1. SLA life cycle metamodel

Fig. 1 depicts a metamodel of the SLA Life Cycle, an improved version of the metamodel presented in [10]. Beside the metaclasses in the diagram, there are also OCL rules underlying the model to enforce static semantics (e.g. the metrics assigned to a service are the same that are assigned to goals of services), not presented due to the lack of space. This metamodel is described in the next paragraphs.

An *organization* generally refers to any division or department of an organization that is either engaged in providing or consuming the service. The term *customer* is reserved for organizations which are consumers of IT services provided by another entity (the service *provider*).

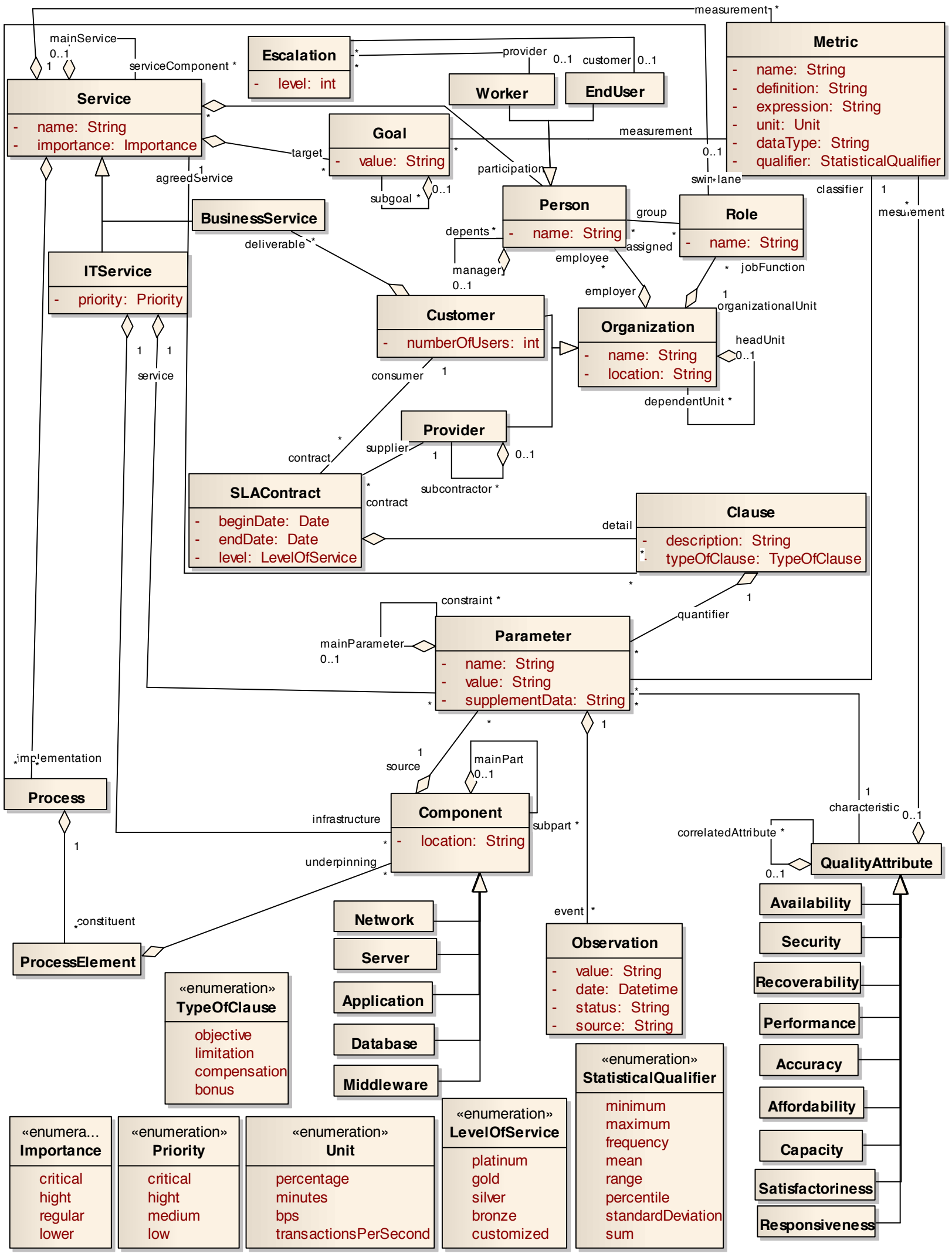

**Fig. 1.** SLAs life cycle metamodel

A customer can be internal or external to the company that provides the IT service. Likewise, service providers can be the own company's internal IT department, or external service providers such as communications service providers (telcos), application service providers (ASPs), internet service providers (ISPs), outsourcing companies or other service providers. The IT department of a company can also be seen as a customer of an external service provider. Therefore, all principles set forth to IT internal customers, will equally apply to IT in its role as a user of external IT services, looking for ways to control costs and achieve consistent service levels.

A customer is the organization responsible for delivering one or more *business services*. Each *service* targets one or more predefined *goals* (with possible *sub-goals*), which is measured using appropriate *metrics* (e.g. number of items dispatched for time unit). A service may need the contribution of other services (depicted by the recursive association in service).

Consumers and providers have their own internal organizational structure. The hierarchical dependency among internal organizational units is depicted as a recursive association in the organization entity. The location of each participant in the SLA contract is relevant when planning where IT services should be provided.

Both customer and provider have *persons* involved in service delivery (mainly *end-users* in the case of business services and IT *workers* for IT services) each one with its specific *role* (e.g. order entry clerk, network operator).

*Business services* and *IT services* are kinds of services realized by mean of processes, whose *elements* (e.g. task, data object) are underpinned by *components* (*network*, *servers*, *applications*, *databases*, and *middleware*) individually considered or brought together in systems.

An *SLA contract*, signed between customer and provider, consists of a set of *clauses*, each one addressing an IT service. For each IT service, a set of *parameters* define the quality attributes' thresholds agreed between customer and provider (e.g., availability in the work period should be 99.99%).

*Quality attributes* are non-functional requirements of IT services that customers expect be fulfilled and providers are compromised to accomplish. As goals in business services, parameters have metrics associated. Some of the types of quality attributes are *availability*, *security*, *recoverability*, and *performance*, each of one has its own metric. The thresholds established by the SLA contract can be measured through observations in order to monitor the degree of accomplishment or possible violations of SLAs.

The components involved in the realization of each IT service are known. In addition, it is also known which IT services contribute for parameters thresholds achievement of each clause of SLA contract. Therefore, one could estimate the contribution of each component for the achievement of agreed services' quality attributes and how component fault can impact those qualities. Ultimately, one can relate the quality of IT services to business value (the recursive association depicted in *service* class relates IT services to business services).

## 2.2. Process notation metamodel

Fig. 2 depicts a BPMN metamodel extract [11]. The full metamodel includes well-formedness rules in OCL (e.g. a message flow can only connect elements in two distinct pools). The BPMN process modeling language was chosen because it is well suited for services representation and allows adding SLAs additional information. Furthermore, it is widely used by practitioners, as well as in the scientific community interested in process modeling [12], since academic search engines (Google Scholar, Microsoft Research and ISI WoK), as well as regular search engines (Google, Bing, and Yahoo), returned almost identical number of hits for “petri net” and BPMN techniques, since 2004, the year of BPMN’s inception.

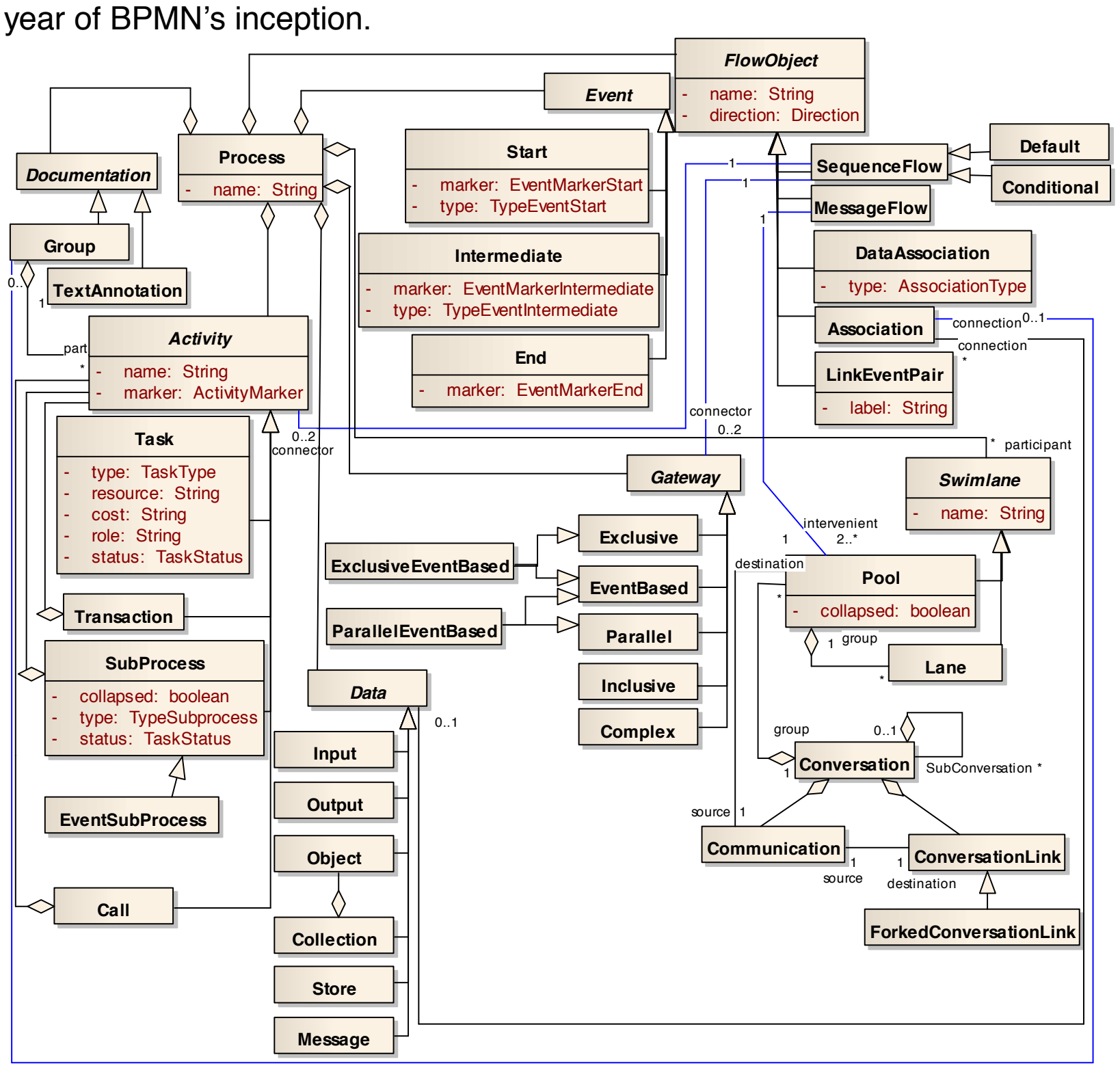

**Fig. 2.** BPMN metamodel

BPMN encompasses five main concepts (*Activity*, *Event*, *Flow Object*, *Swimlane*, *Data Object*), represented as metaclasses, from which all other are specialized. The elicitation of the concepts and relationships of BPMN, allowed the detection of junctions with service level management, as detailed in the next section, in order to build the abstract syntax model of SLALOM.

### 2.3. Metamodel composition

Since the two previous metamodels include modeling concepts abstracting the same real world entities (see Table 1), we used those concepts as junctions to connect and unify the two languages. The metamodel composition technique allows, through the equivalence operator [9], the full union between two UML classes that are converted into a single class. The resulting class includes all attributes and relationships (including associations, generalization, specialization, and containment) from the composed classes.

The resulting abstract syntax will be the hook for semantics to be added to the language specification.

## 3. Concrete syntax models

It is common for languages to have multiple concrete representations: textual, graphical or a combination of both. The abstract syntax model is what unifies the apparently diverse representations, which means that the same abstract syntax model can be presented concretely in either a graphical or a textual format.

In the following sections we introduce some of the concrete syntax models that were considered for SLALOM, addressing aspects such as SLA compliance checking, SLA representation in process models and SLA models validation and verification.

### 3.1. SLA compliance checking

One of the concrete syntax models of SLALOM is used for SLA compliance checking. The model, in textual form, can be fed as input to the USE tool (an OCL specification and validation environment) [13]. This way, it is possible to analyze the model behavior using either actual / diachronic data, collected from system management tools, or simulating different scenarios using the Monte Carlo method for sampling generation. Service level manager and IT staff can now formally check constraints (invariants, pre, post-conditions) against specified thresholds.

After instantiating the model with objects representing the provisioning of resources and consuming of IT services, USE makes possible to query quality attributes of IT services, by evaluating OCL expressions, and discovering possible SLAs non-compliance.

**Table 1.** Common concepts to SLAs and BPMN

| | *Concept* | *SLAs* | *BPMN* |
|---|---|---|---|
| 1. | A role is a set of integrated and coherent activities assigned to an entity (e.g. person, worker, end-user, system or device) inside an organization, which contribute to a global process. A single entity may play several roles and, conversely, a given role may be played by multiple entities. | *Role* | *Lane* |
| 2. | An institution that groups fully differentiated structural and functional units with a common purpose. The organization supplier of IT services is the Provider. The organization that acquires IT services is the Consumer. | *Organization* | *Pool* |
| 3. | Services are the outcome of organization's activities. The outcome of provider's activities are IT services, whereas consumer's produce business services. Services are underpinned by Processes. | *Service* | *Process* |
| 4. | To accomplish a process implementation, a set of elements must be join together, and treated as a unit, for the purpose of process's outcome. | *Process Element* | *Activity, Data, Event, Gateway, Flow Object* |
| 5. | Some system components (applications, middleware, and servers) fulfill specific task in the context of processes that realize IT services. | *Application, Middleware, Server* | *Task* |
| 6. | Network infrastructure is fundamental to allow connectivity among process participants in disparate locations. | *Network* | *Flow Object* |
| 7. | The component database ensures access to actual and diachronic data from process instances. | *Database* | *Data Object* |

**Table 2.** Conceptual Mapping between the SLAs Life Cycle and BPMN Metamodels

| SLAs concept | BPMN implementation |
|---|---|
| Metric | Rule Event |
| SLA Contract, Clause | Process |
| Goal | Rule Event |
| Parameter | Rule Event |
| Quality Attribute | Data Object |
| SLA violation | Signal, Conditional or Timer Event |
| Observation | Data Object |

### 3.2. SLA representation in process models

A concrete syntax model of SLALOM intends to graphically represent SLAs in process models of IT services. Table 2 matches the SLAs concepts with BPMN concepts. We can figure out, for instance, that *Rule Events* contain the rules included in SLA clauses (metrics, goals and parameters), and SLA violations are denoted by throwing signal events, which will be captured and processed according to a workflow defined in the SLA contract (e.g. penalty computation, escalation procedure).

Some BPMN modeling elements can be used to represent SLA concepts, thus leading to an easier inclusion of SLA rules in the IT service process model (as will be explained in Section 4. and depicted in Fig. 3). This is expected to facilitate SLA specification in the design phase of IT service, as well as the interpretation of events during SLA monitoring.

Since this representation intends to target different stakeholders (e.g. end-user, service level manager or system administrator), it is possible by transformation to raise (or lower) the level of abstraction, by release (or include) detail in the model to adapt it to the specific public, although maintaining the validity and coherence of the model.

### 3.3. IT services validation and verification

This concrete syntax model of SLALOM is concerned with the verification and validation of properties of IT services in process models, with support of the ProM tool [14]. In this context, the models used are graphs where nodes represent the concepts in SLALOM and edges represent relationships among these concepts.

Model checking of graphs consists in performing verifications and automatically proving that a property is satisfied. Some common properties of IT service models that can be checked are: *reachability* (some particular situation can be reached), *safety* (under certain conditions, an event never occurs), *liveness* (under certain conditions, some event will ultimately occur), *fairness* (under certain conditions, an event will occur - or will fail to occur - infinitely often), *deadlock-freeness* (the system can never be in a situation in which no progress is possible) [15].

Since SLA violation involves the notion of order in time, temporal logic is used, since this is a form of logic specifically tailored for reasoning with this kind of statements.

## 4. Illustrative example

Due to a space constraint the example introduces only the SLALOM concrete syntax regarding process models, mentioned in section 3.2. We have chosen an ATM example where customers access IT based financial services such as cash withdraw and deposit.

A typical clause of a SLA contract for a set of ATMs, between the financial institution operations department and the IT department, would be: the **service** must be available **99%** of the time from **02:00 to 00:00**, **Monday to Sunday**. Any individual outage in excess of **30 minutes** or sum of outages exceeding **1.5 hours** per month will constitute a violation (see Fig. 3).

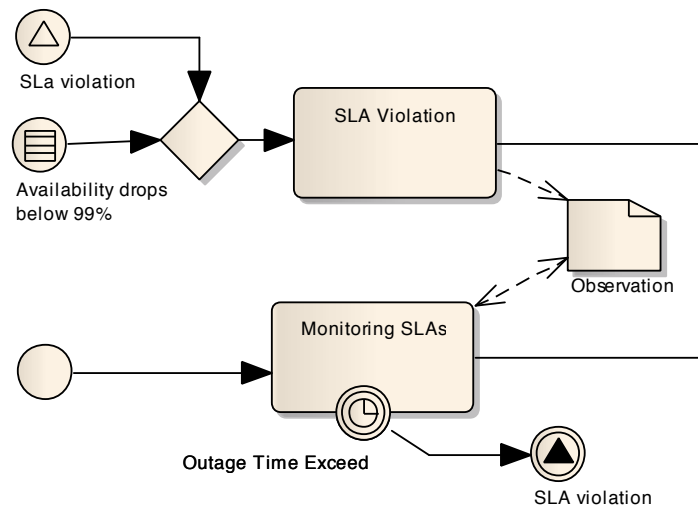

**Fig. 3.** ATM IT service (service level manager perspective)

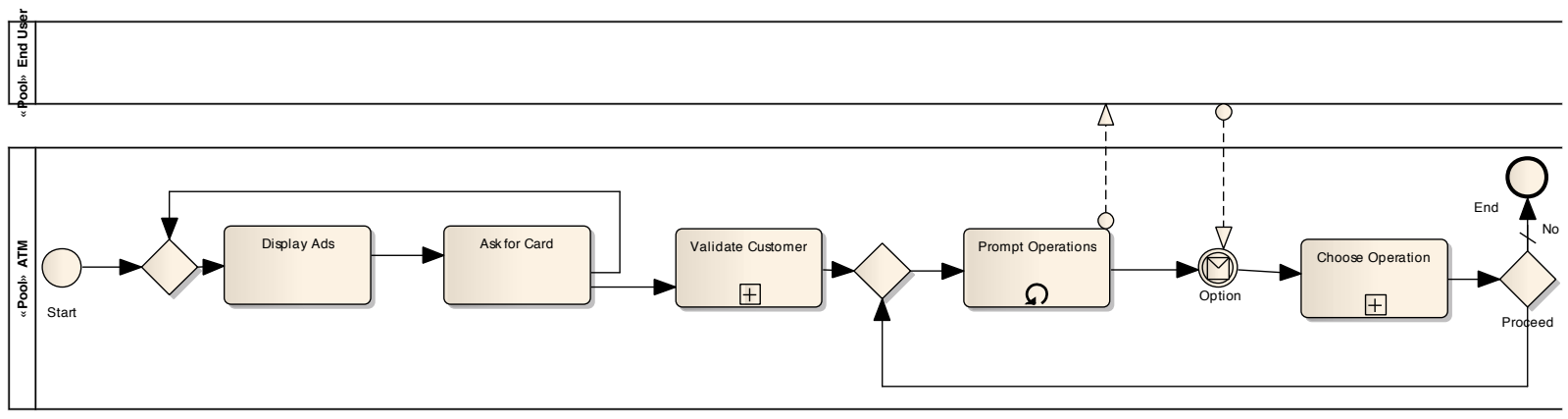

**Fig. 4.** ATM IT service (end-user perspective)

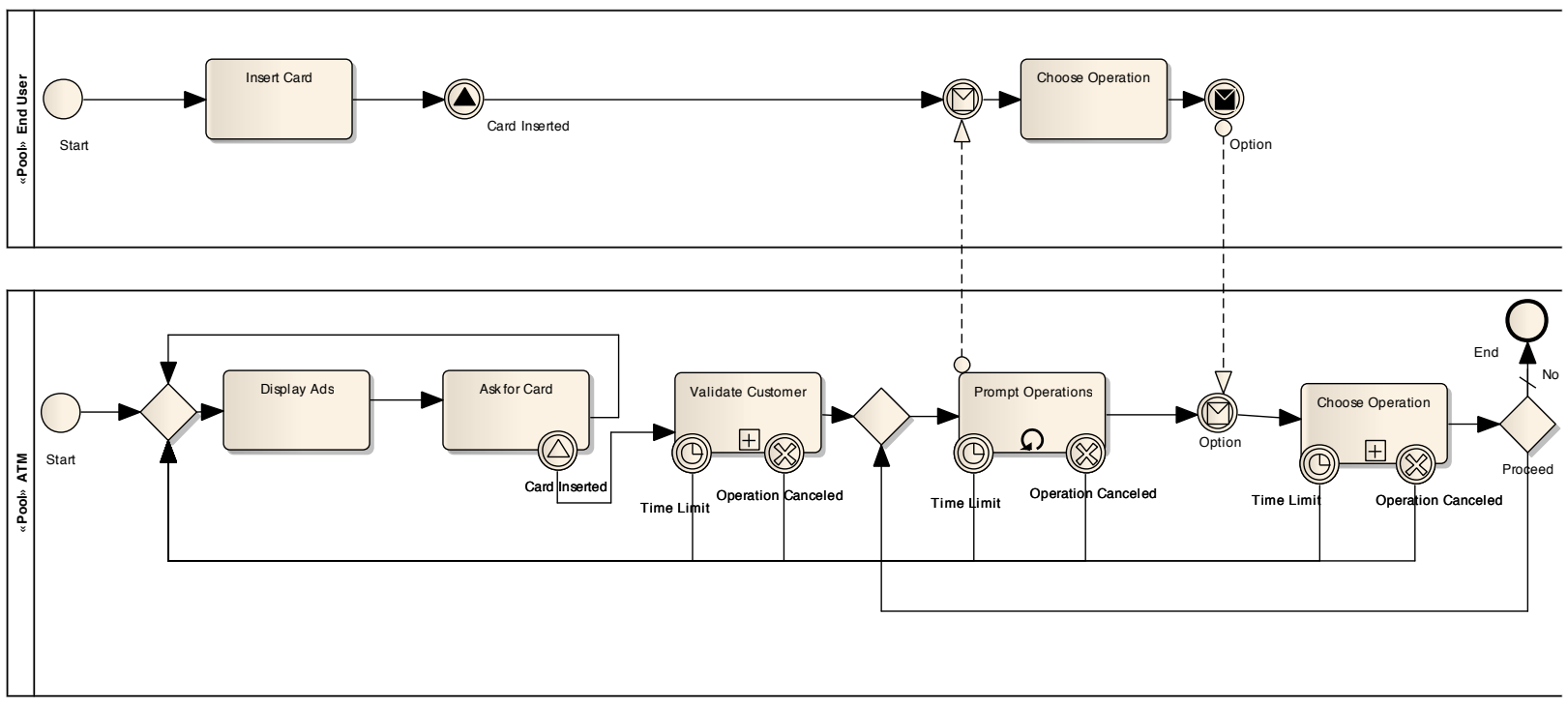

**Fig. 5.** ATM IT service (service level manager perspective)

The concrete syntax model can coherently represent different levels of abstraction: from the end-user perspective (Fig. 4) to a more detailed view of the service level manager (Fig. 5), or even a system administrator view showing interactions with system components or periphericals (e.g. card reader, cash drawer).

## 5. Related work

The field of computer science where SLA specification has gained more attention is in SOAs and in particular web services [16].

One relevant example of a language for SLAs in the context of SOA is SLAng [17]. SLAng was defined by using a metamodeling approach with a high degree of precision in the specification of its semantics, traceability from SLA to language specification, and the testing of the language and SLAs to ensure they capture stakeholders' intents. SLAng supports the expression of mutually-monitorable SLAs, for which the determination of compliance depends only on events visible to both service client and provider. Other examples of SLAs in SOA are, among others, WSLA [18], WS-Agreement [19] and RBSLA [20]. In SOAs and web services the focus are in policies and exchanging messages among machines. However, this is not the context of SLAs in ITSM, due to relevance of human intervention, which must be considered in the models.

## 6. Conclusions and future work

In IT service management, service level agreements (SLAs) are essential to guarantee the quality of provided services. However, there is a gap between the customer business perspective of SLAs and service provider SLAs implementation and monitoring. This paper tries to address this issue, by proposing a domain specific language (SLA Language for specificatiOn and Monitoring - SLALOM). An abstract syntax model of SLALOM was derived based in the composition of metamodels from BPMN and SLA life cycle. The concrete syntax models of SLALOM have different aims, such as SLA compliance checking, SLA representation in process models and models validation and verification. We are planning to build an environment to support those concrete syntax models, including generation tools and tool interoperability.

## Acknowledgments

This work was partly sponsored by the CITI research center at FCT/UNL.

## References

[1] Sturm, R., Morris, W., and Jander, M.: *Foundations of service level management*, Sams, Indianapolis, Ind., 0585309337, 2000.

[2] ITIL3Sm, OGC-Office of Government Commerce: *Summary, ITIL Version 3*, ITSMF- IT Service Management Forum2007.

[3] Correia, A., and Brito e Abreu, F.: '*Model-Driven Service Level Management* ', PhD Research Plan Setptember, 2010, 2010.

[4] Hiles, A.: *The Complete Guide to IT Service Level Agreements - Aligning IT Services to Business Needs*, The Rothstein Catalog on Service Level Books, 1931332134, 2010.

[5] MDA, OMG: '*Model Driven Architecture (MDA)*', No. ormsc/2001-07-01, July 9, 2001, 2001.

[6] Kleppe, A.: *Software Language Engineering: Creating Domain-Specific Languages Using Metamodels*, Addison-Wesley Professional, 1st edn, 0-321-55345-4, 2009.

[7] UML: '*UML-Unified Modeling Language (OMG UML), Infrastructure, V2.1.2*', OMG-Object Management Group, 2007.

[8] OCL, OMG-Object Management Group: 'Object Constraint Language (OCL)', *OMG Available Specification*, 2006, Version 2.0.

[9] Ledeczi, A., Nordstrom, G., Karsai, G., Volgyesi, P., and Maroti, M.: 'On Metamodel Composition', *Proc. of Proceedings of the 2001 IEEE International Conference on Control Applications, 2001. (CCA '01)*, Mexico City, Aug. 06, 2002, 2001.

[10] Freitas, J., Correia, A., and Brito e Abreu, F.: 'An Ontology for IT Services', *Proc. of 13th Conference on Software Engineering and Databases (JISBD'2008)*, Gijón, Spain, 2008.

[11] BPMN, OMG: '*Business Process Model and Notation (BPMN)* ', dtc/2009-08-14, 2009.

[12] Weske, M.: *Business Process Management - Concepts, Languages, Architectures*, Springer-Verlag Berlin Heidelberg, 978-3-540-73521-2, 2007.

[13] Gogolla, M., Buttner, F., and Richters, M.: 'Use: A UML-based specification environment for validating UML and OCL', *Science of Computer Programming*, 2007, pp. 69:27-34.

[14] http://www.processmining.org/prom/start, accessed 30-06-2010.

[15] Berard, B., Bidoit, M., Finkel, A., Laroussinie, F., Petit, A., Petrucci, L., Schnoebelen, P., and McKenzie, P.: *Systems and Software Verification - Model-Checking Techniques and Tools*, Springer, Berlin, 3-540-41523-8, 2001.

[16] Bianco, P., Lewis, G.A., and Merson, P.: '*Service Level Agreements in Service-Oriented Architecture Environments*', No. CMU/SEI-2008-TN-021, Software Engineering Institute - Carnegie Mellon University, 2008.

[17] Skene, J., Lamanna, D., and Emmerich, W.: 'Precise Service Level Agreements', *Proc. of the 26th International Conference on Software Engineering (ICSE'04)*, Edinburgh, Scotland, May 2004, 2004.

[18] IBM: '*WSLA Language Specification, version 1.0*', 2001.

[19] OGF, Alain Andrieux, Czajkowski, K., Dan, A., Keahey, K., Ludwig, H., Nakata, T., Pruyne, J., Rofrano, J., Tuecke, S., and Xu, M.: '*Web Services Agreement Specification (WS-Agreement)*', Open Grid Forum, 2007.

[20] Paschke, A.: 'RBSLA - A Declarative Rule-based Service Level Agreement Language based on RuleML', *Proceedings of the International Conference on Computational Intelligence for Modelling, Control and Automation and International Conference on Intelligent Agents, Web Technologies and Internet Commerce*, 2005, 02, pp. 308 - 314.